\documentclass[12pt,preprint]{aastex}
\pdfoutput=1

\shorttitle{Visualization and  Analysis of Astrophysical Data}
\shortauthors{M. Comparato et al.}

\begin{document}


\title{Visualization, Exploration and Data Analysis of
 Complex Astrophysical Data}


\author{M. Comparato, U. Becciani, A. Costa and B. Larsson}
\affil{INAF-Osservatorio Astrofisico di Catania, Italy}

\author{B. Garilli}
\affil{INAF-Istituto di Astrofisica Spaziale e Fisica Cosmica, Italy}

\author{C. Gheller}
\affil{CINECA - Casalecchio di Reno, Italy}

\and

\author{J. Taylor}
\affil{Institute for Astronomy, University of Edinburgh - United Kingdom}




\begin{abstract}
In this paper we show how advanced visualization tools
can help the researcher in investigating and extracting information
from data. The focus is on VisIVO, a novel open source graphics application,
which blends high performance multidimensional visualization techniques and
up-to-date technologies to cooperate with other applications
and to access remote, distributed data archives.
VisIVO supports the standards defined by the International
Virtual Observatory Alliance in order to make it interoperable
with VO data repositories. The paper describes the basic technical details and features
of the software and it dedicates a large section to show how VisIVO
can be used in several scientific cases.
\end{abstract}




\section{Introduction}

The astronomical community has always dedicated special
attention to the growth of graphical and visualization tools, driving their evolution or even being
directly involved in the development of many of them.


At present, the most popular software for astronomers can be subdivided
into two main categories: tools for image display and processing and tools for
plotting data. Notable among the former are IRAF, by NOAO; ESO-MIDAS, by the 
European Southern Observatory; SaoImage, by the Smithsonian Astrophysical Observatory and
GAIA, by ESO. Many other tools are available, but we refer to dedicated surveys
for a complete list. Gnuplot and SuperMongo are 
popular applications adopted for 2D data plots. A more sophisticated solution 
is represented by IDL, by ITT Visual Information, which is characterized
by a large library of functions specifically developed for astrophysics.
Again, for a complete list, we refer to specific surveys.

Among the most popular N-body visualization codes used by the  
community there are: TIPSY, motivated by the need to quickly display and analyze the  
results of N-body simulations, it is mainly limited to this type of  
data; ParaView, produced by Kitware in conjunction with  the Advanced  
Computing Laboratory at Los Alamos National Laboratory (LANL), the  
goal of the project is to develop scalable parallel processing tools  
with an emphasis on distributed memory implementations; IDL, mentioned above,
contains support for N-body data display, but is not free software.


A new generation of graphic software tools is now emerging. These tools are designed
to overcome the limits
and the barriers of traditional software by exploiting the latest technological 
opportunities. The main challenges and objectives are the following:
\begin{itemize}
\item High performance and multi threading, in order to exploit multi-core systems, large memories and powerful
graphic cards and co-processors. This allows the user to handle large amount of data
in real-time.
\item Interoperability, allowing different applications, each specialized in doing 
different things, to interact with each other in a coordinated and effective
way according to well-defined protocols.  The aim is to provide to the user a complete suite of tools to best analyze  
his/her data. Huge, monolithic and often inefficient tools are obsolete.
\item Collaborative work. The tools allow several users to work on the same data at the same time
from different places, exchanging experience, information and expertise.
\item Access to distributed resources, via web services and/or Grid protocols.
Often, data can no longer be moved from data centres as it is too large and complex. The astronomer 
must have the tools to access it, independently of his geographical location in a fast 
and reliable way.
\end{itemize}
Tools like VisIVO, Aladin \citep{bon00}
and Topcat \citep{Topcat}, have been recently developed in the 
framework of the Virtual Observatory (www.ivoa.net) to achieve all or some of these 
goals. In this paper we will focus in particular on VisIVO, which stands for Visualization Interface
for the Virtual Observatory. VisIVO is being developed as
a collaboration between the Italian National Institute for Astrophysiscs (INAF) - Astrophysical Observatory 
of Catania and CINECA (the largest italian academic high performance computing centre)
in the framework of the FP6 EU funded VO-Tech project. The next section gives a short review of the basic
functionality of VisIVO, while section 3 describes
PLASTIC, a messaging protocol which allows heterogeneous applications 
to work together. Part of section 3 and section 4 are dedicated to presenting several scientific cases in which the support of graphics 
and visualization is of primary importance. In these sections we show some of VisIVO's 
capabilities in action, demonstrating how they can be effectively used in  practical
applications.

\section{VisIVO}

VisIVO is a C++ application specifically designed to deal with multidimensional data.
It is free software available both for MS Windows and for GNU/Linux (porting to MacOS
is in progress). It can be downloaded from the web site http://visivo.cineca.it.
The software is built on the top of the Multimod Application Framework (MAF)\citep{vic04}. 
MAF is an open source framework for the development of data visualization and analysis applications. 
It provides high level components that can be easily combined to develop a vertical 
application. It is being developed by the 
visualization group of CINECA  and it can be downloaded from the web site http://openmaf.cineca.it.
The framework is based on the Visualization ToolKit (VTK) \citep{sch04} library for
the multidimensional visualization and on the wxWidgets library, a
portable Open Source GUI library, for  the user interface. It incorporates other open libraries, 
(for example for data encryption) or drivers for virtual reality devices (3D mouse, gloves, haptics etc.).
VisIVO's architecture strictly reflects the structure of a typical
scientific application built with the MAF, being mainly
developed in the highest layers of the framework.
The software exploits, wherever possible, the standard visualization services,
views, operations and interface structures provided by the framework and
implements all the elements that are specific to the visualization and
analysis of astronomical data.

Extensions to the basic MAF infrastructure have been developed in order to match astronomy-specific 
requirements and to provide the highest performance. Internal data representation is in the form of a {\it Table Data} structure, 
which is composed of a sequence of variables loaded from a data source such as a file or a database. 
Regardless their original type, variables are all converted to {\it double} format. While this incurs a penalty in the 
application's memory needs, it provides  the necessary precision in some of the data processing stages. 
Once a table is loaded the user can manage and visualize the data. These operations do not increase the 
memory usage as long as they do not create new tables or new fields: 
the visualization process is carried out using references to the
Table Data with no data replication. In order to visualize data, the user has to set which of the 
loaded fields will be used as the coordinate system of a Cartesian reference frame. In this way, the 
software ensures maximum flexibility in data usage.

\subsection{VisIVO for data visualization} 

Data visualization is the main target of VisIVO. The software is designed to simultaneously handle 
as many properties as possible. Complex tables can be loaded and manipulated, new fields can be 
derived and finally represented graphically, using points, colours, transparencies, surfaces, glyphs and
volume rendering.
The first step of a working session is usually data loading. Data can be read from files;
VisIVO supports different kinds of file formats: standard file formats, like  VOTables, FITS, HDF5, ASCII, 
raw binaries and the native data format of the popular $Gadget$ simulation code \citep{spr00}.
The VOTable format is an XML standard for the interchange of astronomical data, defined by
the International Virtual Observatory Alliance (IVOA, http://www.ivoa.net). 
Data is represented
as a set of tables, each table being an unordered set of rows, 
whose format is specified in the table XML metadata. Rows are
sequences of table cells, each containing either a primitive data type
or an array of such primitives. VOTables can also contain links to external files
as a separate data source. VisIVO uses the Savot VOTable parser developed by CDS\footnote{Centre de Donnes astronomiques de Strasbourg} to load and write VOTables. 
FITS and HDF5 importers are implemented using the published API and libraries.
The ASCII table format consists of columns of data spaced with the most common separation characters (space, tab etc.). 
Raw files are sequences of variables written as binary dumps of the memory. The binary files can be managed by 
descriptor files which store the associated information (number of variables, data types etc.). 
VisIVO can also interact with CDS VizieR
data service \citep{och00}, retrieving data directly from remote archives (see par. 2.3).
\begin{figure}
\begin{center}
\includegraphics[width=11.0cm]{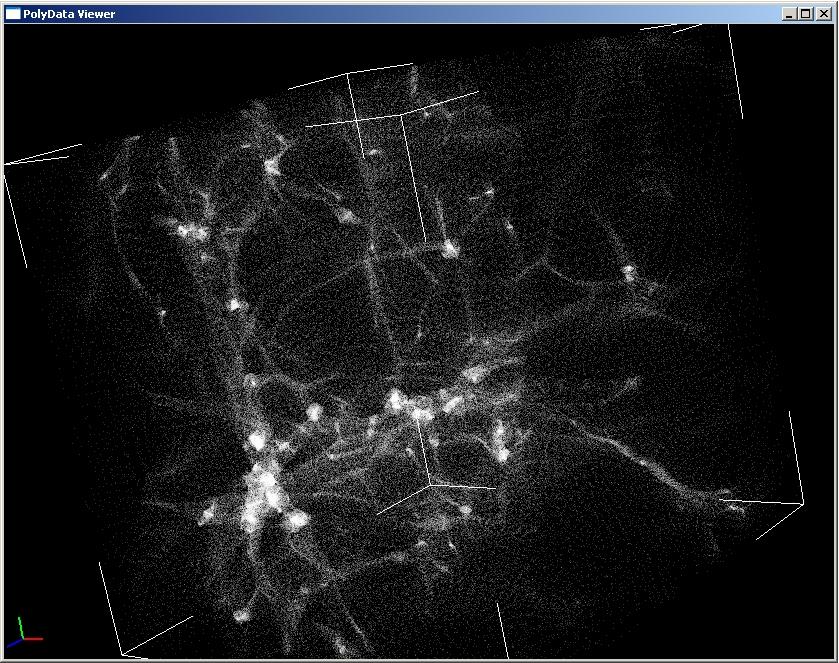}
\includegraphics[width=11.0cm]{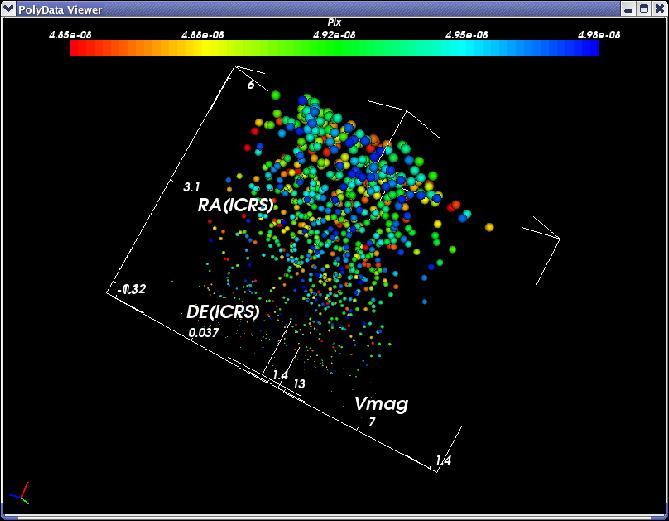}
\caption{3D data display of 16 million-particles N-Body simulation and data from VizieR Astronomical 
Server cone search: the Hipparcos and Tycho Catalogues (ESA 1997)}
\label{fg1}
\end{center}
\end{figure}

Once data is loaded it can be visualized and analyzed. VisIVO can deal with both structured and unstructured
data. The former is represented by fields defined on a regular mesh. The latter is data with no 
special geometry; it is treated as sets of points. 
Graphically, unstructured data have a default representation as pixels. The points geometrical distribution is set  
by selecting three of the fields loaded as Table Data. For example: the
$x$, $y$, $z$ coordinates of the particles from a N-Body simulation, or 
the RA (Right Ascension) - DE (Declination) - Vmag (Johnson magnitude V)
fields for data from stellar catalogues (figure \ref{fg1}).

Besides their geometric position, points can be used to display further quantities, using 
colours and glyphs (3D shapes, like spheres or cubes). Points can be coloured as a function 
of a given scalar field (e.g. their temperature or their spectral index) with a colourmap that the user can customize. 
Each point can also have an associated glyph, whose size 
can be a function of one (for spheres) or two (for cubes, cylinders, pyramids) fields. A vector quantity
can be visualized as well, using either oriented segments or arrows. Vectors can also be coloured according
to their magnitude.

\begin{figure}
\begin{center}
\includegraphics[width=11.0cm]{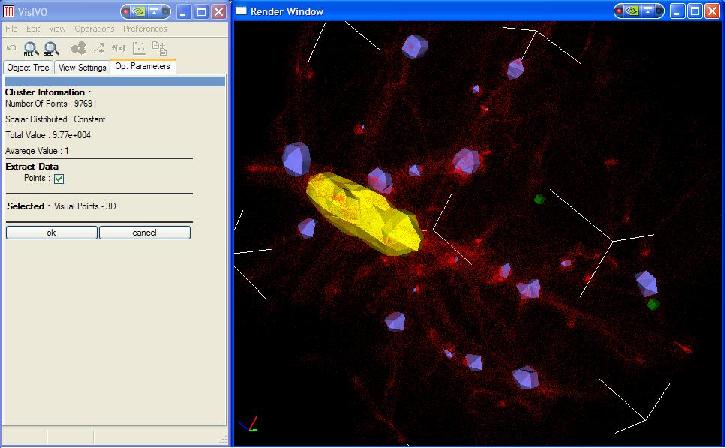}
\caption{Clusters identification in data from a cosmological simulations. 
Points inside the yellow isosurface can be extracted.}
\label{fg2}
\end{center}
\end{figure}

If the data size is too large
to be managed in memory, VisIVO allows the user also to extract a random subset of points.
It is also possible to select the points which lie in a region the user is specifically
interested in (e.g a galaxy cluster in a cosmological simulation, or a globular cluster in a
catalogue of stars). The selection can be accomplished using either a rectangular 
sampler or the cluster finder utilities. For the latter, the following cluster identification method is
implemented. 
A field associated to the points (e.g. the point mass) is used to set a threshold. All the points that have 
a value of the field above the threshold, comprise a cluster. Surfaces which divide regions
above and below the threshold can be visualized (see figure \ref{fg2}). Regions geometrically disjoint 
(i.e. their threshold surfaces do not overlap) are identified as 
separate clusters.

Structured mesh-based data do not have a default graphical representation. VisIVO can visualize them 
using volume rendering and isosurface.
Volume rendering is a visualization technique in which the field values
are represented by different colours and different
transparencies. The global effect is a cloud appearance. This method enables the user
to emphasize also the inner parts of the volume. 
Isosurfaces  are surfaces of given value calculated from a mesh-based quantity. 
The isosurface can be defined as the surface which 
divides regions in which a given field has value above a certain value 
from regions in which it is below that value.

\subsection{VisIVO for data analysis} 

VisIVO provides various built-in utilities that allow the user to perform 
mathematical operations and to analyze data. 
It is possible to apply algebraic and mathematical operators to the loaded data. Basic arithmetic 
operations (addition, subtraction, multiplication, division) as well as logarithm, power law, absolute
value and many others are supported. Scalar product, magnitude and norm of vector quantities 
are available too. 
In this way, new physical quantities can be calculated. For example, for the gas distribution 
of a simulated galaxy cluster, the X-ray emission due to 
thermal Bremsstrahlung can be calculated as proportional to the product of the 
square of the mass density and the square 
root of the temperature of the gas. If these two quantities
are available, the emission can be immediately derived.
It is also possible to merge two different Table Data structures to create a new one. Data in the resulting
table can be treated as a single dataset.
The merging capabilities and the mathematical operations give great flexibility in data analysis and representation.\\
Several built-in functions allow the user to perform a statistical analysis of a points distribution.\\
The {\bf Scalar Distribution} function calculates the distribution of any
quantity loaded in the Table Data and plots it as a histogram.\\
The {\bf Correlation Filter} calculates the linear two-point correlation
function of a point set. 
This is defined as the probability $\delta P$ of finding a point in a randomly-chosen volume $\delta V_{1}$ and a point in
another volume $\delta V_{2}$ separated by a distance $r$.
The two-point Correlation Function of VisIVO is based on the 3D counterpart of the {\em Peebles \& Hauser}
estimator \citep{pee74}:
\begin{equation}
\xi_{PH} = \frac{DD(r)}{RR(r)} (\frac{N_{rd}}{N}) - 1,
\end{equation}
where $N_{rd}$ is the number of points in an auxiliary random sample, $DD(r)$ is
the number of all pairs of points with separation inside the interval $[r-dr/2, r+dr/2]$ and
$RR(r)$ is the number of pairs between the data and the random sample with
separation in the same interval. The random
sample must have a density  $2.5$ times the density of the real dataset.
The  box is divided into
a number $N_{bin}$ of cubic subintervals. Then a frequency histogram
of the pair distances of particles is constructed.
The calculation of $RR$ and $DD$ is performed with a Monte Carlo integration.\\
The Fourier transform of the correlation function is represented by the {\bf Power Spectrum},
which can be estimated by VisIVO as well. 
The power spectrum of a set of N massive particles can be calculated as 
\begin{equation}
P({\bf k}) = \langle \mid \rho({\bf k}) \mid^{2} \rangle.
\end{equation}
where $P({\bf k})$ is the power spectrum, ${\bf k}$ is the three dimensional wave
number: $k_i = 1/r_i$, with $r_i$ indicating the i-th component of the spatial position of a point and
$\rho({\bf k})$ is the Fourier transform of the mass density field.
The power spectrum provides the same statistical information of the correlation 
function, but it is much faster to compute. However, in the present implementation,
periodic boundary conditions are required. Furthermore, in general, the spatial resolution is worse 
than that of the correlation function. In fact, in order to calculate the Power Spectrum, a Cloud in Cells  
algorithm  distributes a constant value for each point (its mass) on a regular structured mesh with
periodic boundary conditions. The mesh resolution sets the maximum wave number that the power spectrum
can be calculated on. With this procedure, a mesh based mass density distribution is reconstructed and
a fast FFT based approach  can be used to estimate $P({\bf k})$.\\
The last available analysis tool is for 
 {\bf Minkowski Functionals} (\emph{MFs}).
They describe the Geometry, the Curvature and the Topology of a point-set \citep{pla95}.
In a three-dimensional Euclidean space, these functionals have a direct geometric
interpretation.  The first Functional represents the volume $V$ of
a structure, the second one represents the surface area $A$ and it is a measure of the geometry of the distribution.
The third Functional corresponds to
the integral mean curvature $H$ of the structure's surface. It represents a measure of the 
distribution topology.
The \emph{MFs} algorithm implemented in VisIVO associates a 
\textit{covering sphere} of radius $r$ to each data point. The size, the shape and the connectivity of the spatial pattern, composed by the union-set of the spheres, change with the radius, which can be used as a diagnostic parameter. In particular, VisIVO computes
the reduced values of the Minkowski Functionals, $\Phi(\mu)$ with $\mu = 0, 1, 2$, that are
the ratio of the \emph{MFs} of the actual distribution to the \emph{MFs} of the same number
of disjointed convex bodies. Their values always start from unity: for small radii, all the   covering spheres are  disjointed.
In the 3rd functional $\Phi_{3}(r)$, the first zero provides an estimate of
the percolation threshold. A spongy structure, like
a Poisson distribution, gives lower values for $\Phi_{3}(r)$,
while higher values indicate structures with few  big filaments or tunnels. 

\subsection{VisIVO and VizieR}

In the age of the Virtual Observatory,
data collections are distributed between various sites.
They are accessible to user applications via standard technologies such
as the Web Service WSDL/SOAP protocol. One of the services exploiting 
this Web Service technology is VizieR, version 2 of which is available on the
CDS servers.  Although it is still in beta, the service will also be available at ADS,
ADAC and CADC after the final release.
VizieR is a database which archives, in an homogeneous way, thousands of
astronomical catalogues gathered over decades by the CDS and participating institutes.
The new web service interface gives access to the VizieR
database of astronomical catalogues by adding four new methods to the old interface:
\begin{itemize}
\item coneCatalogs
\item coneResults
\item ADQLrequest
\item getAvailability
\end{itemize}

VisIVO, using the Axis C/C++ library, implements an interface to the
service. It is able to get the list of available servers using the
getAvailability method, to get the list of valid parameters values to pass
to the coneCatalogs and coneResults and using the last two methods to
get metadata and data (in VOTable format) about catalogues depending on
the given parameters.\\
In this way, VisIVO is able to query directly the VizieR web service to
retrieve data from it and visualise them as if they were local data.
The interaction with the service is transparent to the user. The user
need only fill in specified fields with the parameters defining the
data he wants to download. The result of this operation is a list
of catalogues and, on selecting one of them, data can be visualised as if
they were in a file or saved on the disk in the VOTable format.

\subsection{VisIVO performance}
VisIVO's recommended system requirements are such that it can be used
on a consumer laptop personal computer. VisIVO's performance is  
mainly constrained by the system RAM size; each loaded file  will  
create a sequence of float arrays: the more memory the user has
on his system the more data he can load. On
a laptop  with 1GB of RAM running a GNU/Linux system, VisIVO is able to load and handle
2 million points interactively, while up to 16 million points can be loaded in about 5 seconds, although the  
visualization then becomes quite slow.
On Microsoft Windows XP, VisIVO is limited to 8 million points due to  Windows' memory management.

Table 1 shows the execution time in seconds of typical operations the  user performs with VisIVO.
The test were carried out both with   GNU/Linux systems  and  Microsoft Windows XP (the column names containing GL and MW  letters  respectively) and ATI Mobility Radeon X600  with 64 MB
and ATI Radeon Xpress 1100 graphics cards.
The  16M and 8M letters identify tests with sixteen and eight million  
row datasets respectively. HPS identifies our high
performance dual processor system equipped with two AMD Opteron Dual  
Core 280 (2.4GHz), 8GB of RAM and a
7200rpm SATA hard disk. LAP1 and LAP2 identify two consumer laptops,  
LAP1 is equipped with an Intel Pentium
M 740 (1.73GHz), LAP2 is equipped with an AMD Turion TL50x2 (1.6GHz),  
both have 1GB of RAM and 5400rpm
hard disks. The IMPORT test consists of loading a binary file  
containing six fields,
the GEOMETRY test consists of defining a 3D point distribution from  
three of the loaded fields, the DISPLAY test consists of visualizing  
the geometry and the DISTRIBUTE test consists of distributing a  
scalar value associated to the sixteen million points on a regular  
grid with $128^3$  cells. The  POWER SPEC. test  computes the power  
spectrum of the point distribution and the CORRELATION test computes  
the correlation function of the point distribution. The SUBSAMPLE  
test performs a geometrical subsample of the point distribution, and  
finally the EXT. CLUSTER test  performs an extraction of the points  
within an isosurface of one of the scalar associated with the points.

\begin{deluxetable}{lccccc}
\tabletypesize{\scriptsize}
\tablecaption{VisIVO Performance Test\label{tbl-1}}
\tablewidth{0pt}
\tablehead{
\colhead{TEST} & \colhead{HPS-GL-16M} & \colhead{LAP1-GL-16M} &
\colhead{LAP2-GL-16M} & \colhead{LAP1-MW-8M} & \colhead{LAP2-MW-8M}
}
\startdata
IMPORT       & 5   & 4   & 16  & 7  & 5  \\ 
GEOMETRY     & 18  & 20  & 35  & 6  & 7  \\ 
DISPLAY      & 44  & 27  & 30  & 25 & 21 \\
DISTRIBUTE   & 8   & 9   & 11  & 10 & 5  \\
POWER SPEC.  & 13  & 20  & 16  & 8  & 8  \\
CORRELATION  & 18  & 30  & 25  & 16 & 14 \\
SUBSAMPLE   & 4   & 5   & 24  & 2  & 4  \\
EXT. CLUSTER & 310 & 354 & 420 & 44 & 60 \\
\enddata
\end{deluxetable}

\section{Application interoperability through PLASTIC}

The capabilities of VisIVO are extendable through an application interoperability protocol called PLASTIC \citep{tay06} 
(PLatform for AStronomy Tool InterConnection).  Equally, through PLASTIC, VisIVO's functionality is
made available to other applications.\\  
The motivation for PLASTIC is the desire to leverage the abilities of different desktop applications in a seamless way.  
Scientific applications such as VisIVO are being continuously enhanced with new features and while this is of great benefit to users, 
there are some limitations to this ''bigger is better'' approach.  
Inevitably there will be some overlap between the applications' 
feature sets as users demand capabilities from other applications in their own favourite tool, leading to a duplication of effort
within the scientific software community.  As applications become more powerful their resource 
footprint usually expands, and their 
increased complexity may make them more difficult to maintain and to use.\\
The alternative and complementary approach is to encourage collaboration between applications,
each a specialist in a particular task.  This approach enables the user to assemble a suite of
tools according to his personal requirements.  Exporting and outsourcing functionality to other
applications has been explored by CDS:
Aladin \citep{bon00} exposes a publicly documented interface (''VOApp'') that makes it possible 
for third-parties to write plug-ins \citep{Aladinplugin} that expand Aladin's capabilities and reuse its
functionality.  However, Aladin's plug-ins
are restricted to being Java applications that can run inside the same JVM, which sometimes
leads to resource, packaging and class-compatibility
problems.  The VisIVO and Aladin developers overcame this constraint by making VisIVO, a C++
application, control Aladin through
the latter's scripting interface.  However, the architecture is no longer symmetrical and Aladin is
unable to control VisIVO in return. It was recognized that there was a need to generalize the
VisIVO/Aladin interoperability to arbitrary applications and thus the VOTech 
consortium \citep{VOTech}, of which
the VisIVO team is a founder member, created the PLASTIC interoperability protocol. Through
PLASTIC, applications can share data and link views.  Data exploration using disparate linked
views of the
 data is not a new idea: for example, the Mirage \citep{Mirage} and xmdvtool \citep{Xmdv} 
applications
each supports several visualization methods allowing the user
  to explore data simultaneously using different methods.  PLASTIC extends this concept to allow linked
views across applications.\\
PLASTIC works through a locally running daemon application called a ''Plastic Hub'' and is derived from  
the technology developed for the Astro Runtime \citep{win06}.  
Applications communicate with the Plastic Hub using one of several protocols: 
different protocols are supported to make it easier to adopt PLASTIC by application developers.  
PLASTIC does not define a fixed API of operations that all applications must support.  
Instead, it employs a simple inter-application messaging system: applications send each 
other messages requesting certain actions.  These are sent via the Plastic Hub, which then routes them to their 
destinations.   The current set of messages includes ''load this table'', and 
''select/highlight these data points'', but can be extended by application developers as new 
ways of collaborating arise.  
The advantage of the Hub-based architecture is that individual applications need 
only understand one of the Hub's communication protocols, and the Hub is responsible 
for any required translation.   Furthermore, applications can dynamically discover other 
applications and their capabilities by interrogating the Hub.  
PLASTIC is platform and language neutral and, at the time of writing, has been incorporated into 
more than a dozen applications written in Java, C++, Python, 
Perl, Tcl and JavaScript/html including Topcat, Aladin, VisIVO, VOSpec \citep{VOSpec}, AstroWeka and Reflex \citep{Reflex}.

\subsection{PLASTIC interoperability in action}

\begin{figure}
\begin{center}
\includegraphics[width=11.0cm]{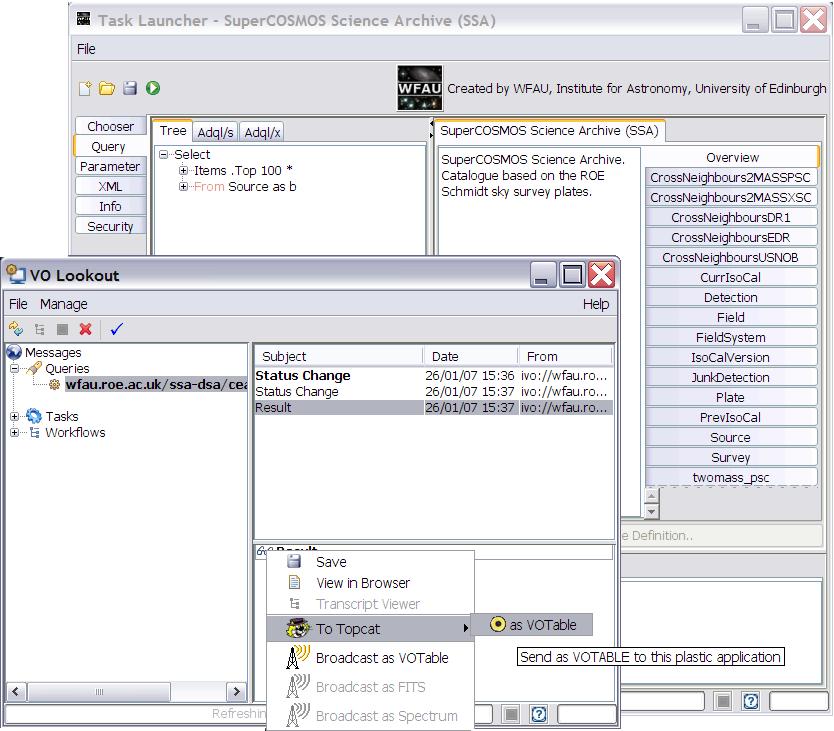}
\caption{Sending the SUPERCOSMOS data from the Workbench to Topcat}
\label{plasticfig1}
\end{center}
\end{figure}

\begin{figure}
\begin{center}
\includegraphics[width=11.0cm]{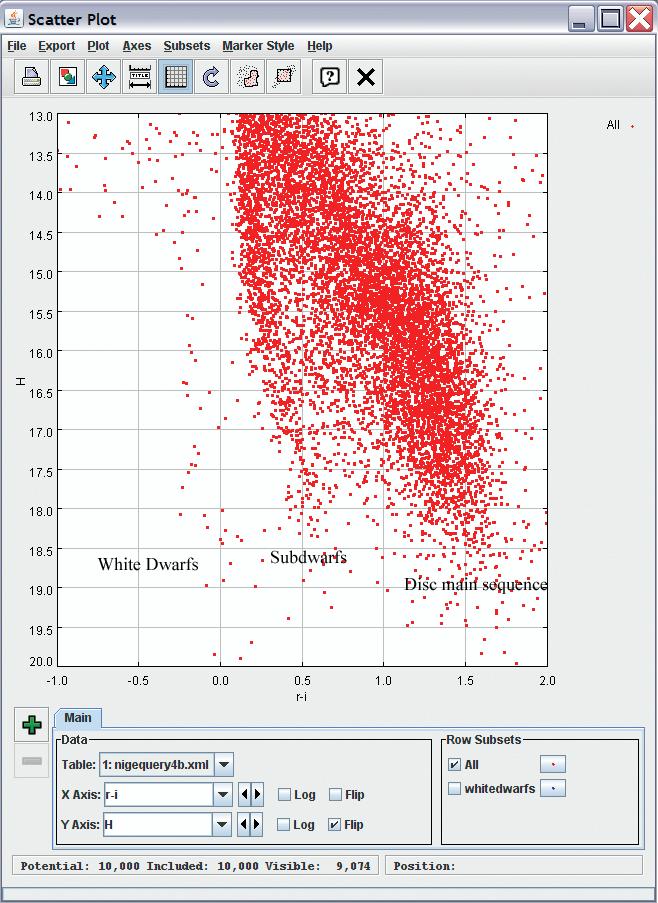}
\caption{The three populations, as seen in Topcat}
\label{plasticfig2}
\end{center}
\end{figure}

The following example illustrates how VisIVO's PLASTIC interaction works in practice, with several applications exploiting each other's 
strengths to explore data taken from the paper by Digby et al. \citep{dig03}.  The AstroGrid Workbench 
\citep{Workbench} is first used to search for suitable data in the SuperCOSMOS 
Science Archive \citep{SSA} and Sloan Digital Sky Survey \citep{SDSS}.  The astronomer begins his analysis 
by starting Topcat \citep{Topcat}, which connects 
to the Plastic Hub.  The astronomer then uses PLASTIC to send the 
SuperCOSMOS data directly to Topcat (figure \ref{plasticfig1}).
The SuperCOSMOS data does not contain the attributes that the astronomer needs, so he uses Topcat to add synthetic columns to calculate the 
colour indices of the sources and their reduced proper motions.  A scatterplot of the r-i colour index against the reduced proper motion 
reveals that the sources fall into three populations: white dwarfs, sub dwarfs and main sequence stars (figure \ref{plasticfig2}).

\begin{figure}
\begin{center}
\includegraphics[width=11.0cm]{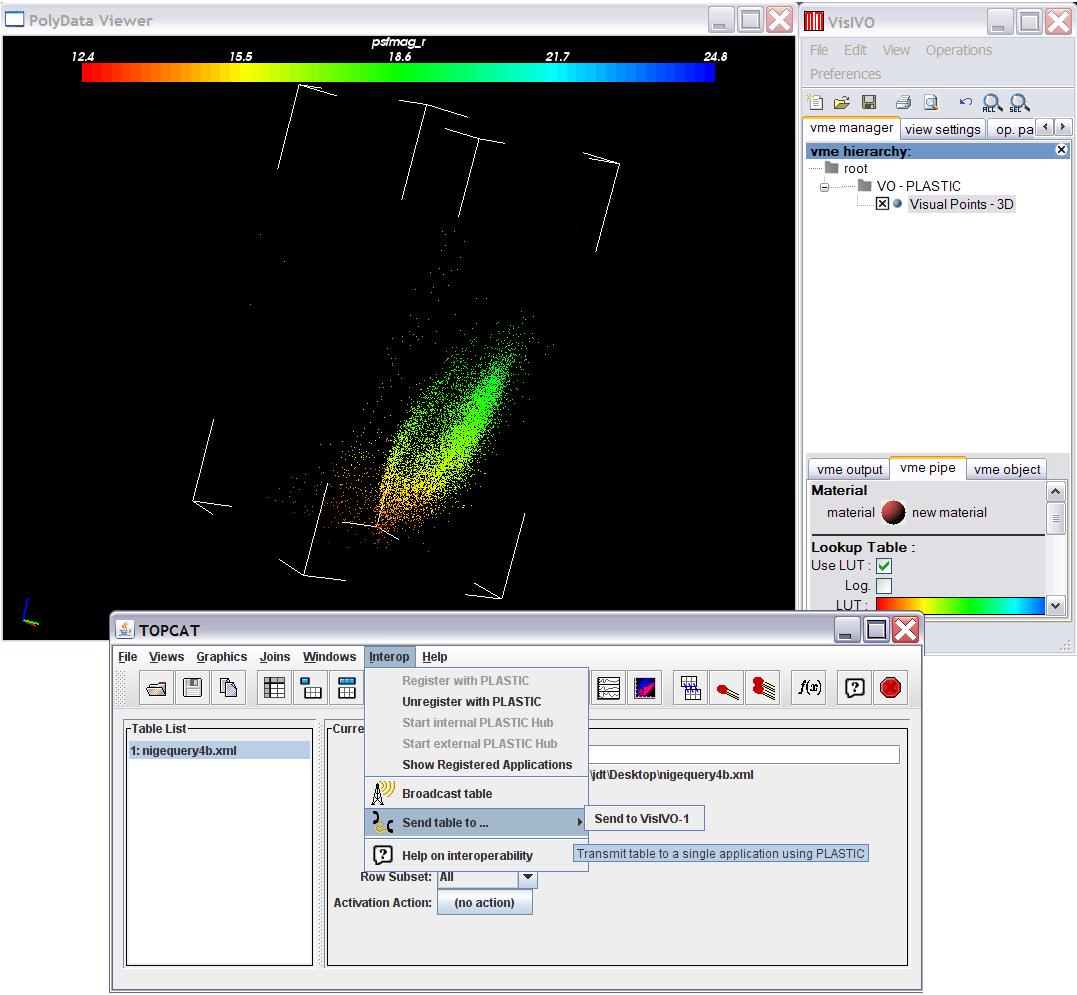}
\caption{Sending the augmented data to VisIVO}
\label{plasticfig3}
\end{center}
\end{figure}

To fully understand this dataset the astronomer starts VisIVO, since it specializes in plotting multidimensional data.  VisIVO 
connects with the Plastic Hub, and shortly afterwards the Workbench and Topcat's menus update to reflect the fact 
that VisIVO is running and able to receive VOTables.    The astronomer sends the augmented data from Topcat to VisIVO and uses VisIVO to 
create a 4D plot of r-i colour index, g-r colour index, reduced proper motion, 
and magnitude (using the latter to choose the colour of the points from a lookup table) (figure \ref{plasticfig3}).

\begin{figure}
\begin{center}
\includegraphics[width=11.0cm]{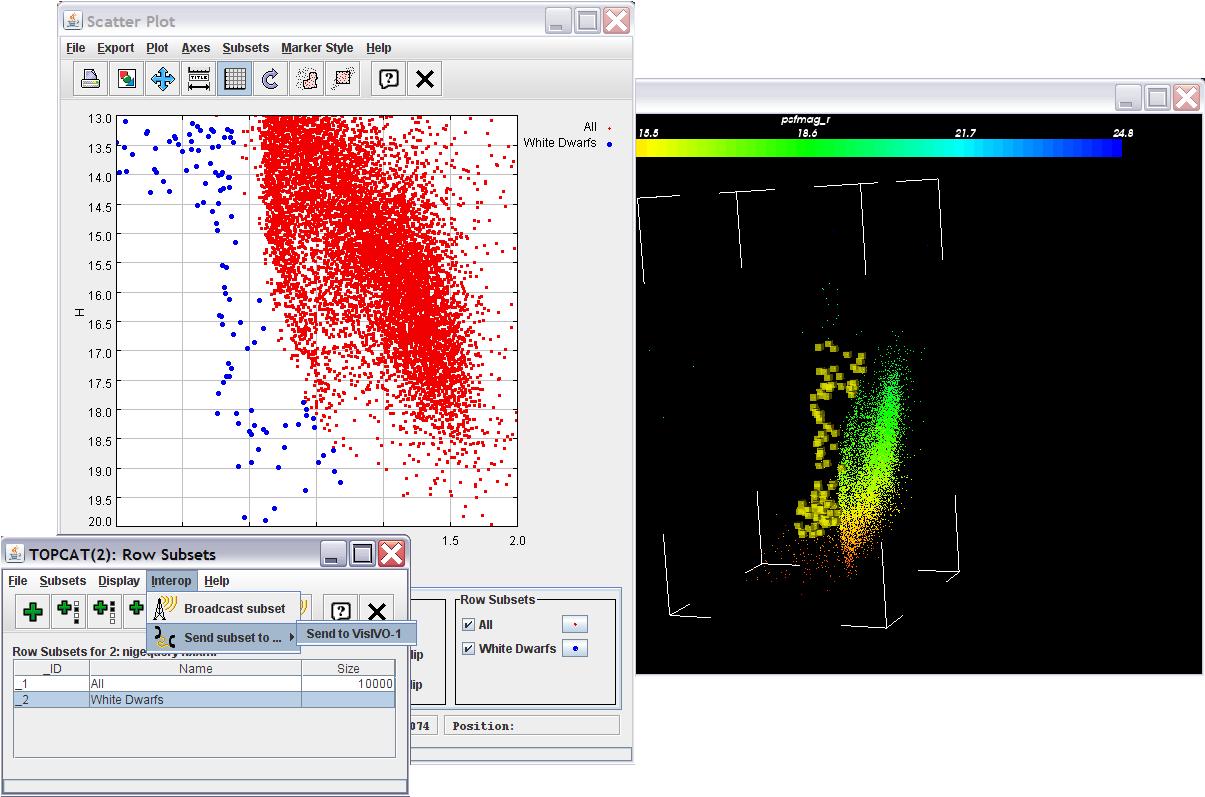}
\caption{Linked views in Topcat and VisIVO}
\label{plasticfig4}
\end{center}
\end{figure}

The white dwarf population can then be selected in Topcat and is automatically highlighted in VisIVO through PLASTIC 
for further exploration (figure \ref{plasticfig4}).   Finally, 
interesting objects can selected using VisIVO's picker tool and sent via PLASTIC to Aladin 
for overlaying over an image so that the astronomer can see their spatial distribution.
This workflow is, of course, greatly simplified.  
In reality the data would be transferred back and forth between VisIVO and other applications, 
with interesting clusters extracted and spurious data removed.  
It could be even be sent to statistical applications such as AstroWeka \citep{AstroWeka} and 
Eirik \citep{Eirik} to aid the astronomer in identifying clusters and trends.

While some of the above workflow could be accomplished by saving the dataset from one application and loading it in the next, 
PLASTIC makes the operation seamless.

\section{Scientific cases} 

In this section, as an example of the practical usage of VisIVO, we will show how the tool can help the researcher in the 
analysis of two different problems: 
the classification of 
galaxies between star forming and quiescent objects and the detection of shock waves in galaxy clusters

\subsection{Galaxies Classification}
\begin{figure}
\begin{center}
\includegraphics[width=11.0cm]{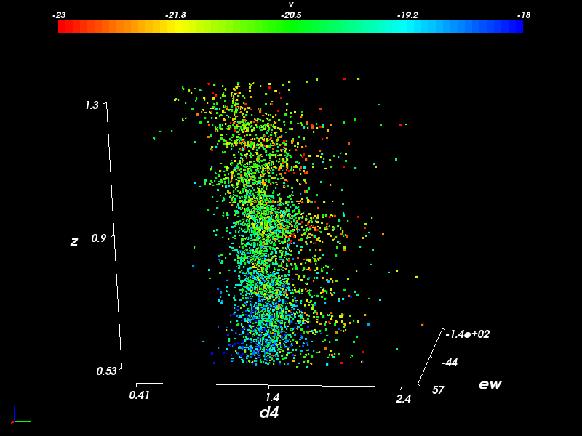}
\includegraphics[width=11.0cm]{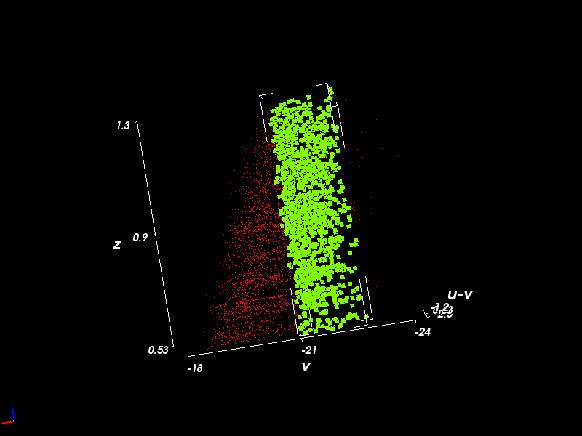}
\caption{Visualization of the sample. On the top (a), the global sample
  where colour indicates the Absolute magnitude. At
  higher redshift, fainter (i.e. blue) galaxies are not visible in the
  sample. On the bottom (b), the selection of a volume limited
  sub-sample
using the absolute magnitude vs. redshift projection of the data cube.
  The green cubes are the selected subsample}
\label{tot_sample}
\end{center}
\end{figure}

One of the most widely used methods relies on the intrinsic-colour
versus absolute magnitude diagram: in this representations, redder objects,
assumed to be quiescent, occupy a well defined locus, also known as
{\it red sequence}. On the other hand, having spectroscopic information
at hand, the equivalent width of the [OII] emission line is a well
known tracer of on-going star formation activity, while the depth of
the break at 4000 $\AA$ is an indicator of stellar age (the higher the
break, the older the stellar population is). The question we ask
ourselves is whether using one or the other of these indicators, we
select the same population of galaxies, and also whether these
indicators show some evolution with redshift.
Starting from the VVDS epoch1 survey 
\citep{lef05}, we have extracted a subsample of galaxies with secure
redshift, for which the
equivalent width of the [OII] emission line 
(simply named [OII] from here on), and the depth of the 4000
$\AA$ break (from here on D4000) has been measured with good
confidence (see Franzetti 2007 for details).
Our global initial sample consists of 4640 galaxies, and for each
object we have a measure of: V absolute magnitude, 
intrinsic U-V colour, redshift,
[OII] and D4000.
As a first step, we import the ASCII file where such information is stored
into VisIVO, and do a first rough visualization of D4000 vs. [OII]
vs. z. Such plot shows that there are $\sim$ 25 objects for which 
an anomalously large [OII] and/or D4000 has been measured. 
As we are currently interested in the global properties of the
sample, and not in the outliers, we use the SubSample function to
select the bulk of our data, eliminating such outliers.
In figure \ref{tot_sample}a, the 3D
representation of the D4000 vs. [OII] vs. z for the remaining objects is
shown. The colour scale
indicates the absolute magnitude associated to each object.

Two effects are immediately visible from  figure \ref{tot_sample}a: the
first one is the {\it to be expected} selection effect on galaxy luminosity: the
further we go in redshift, the more luminous are the galaxies we have in
our sample (we have lesser and lesser blue points as redshift
increases). The second effect we see is a tendency
for D4000 to decrease with increasing redshift. Before further
inspecting this decrease, which can be scientifically very promising,
we get rid of the luminosity selection effect, by extracting from our global sample a volume limited
sub-sample, i.e. we include only objects having an absolute magnitude
visible throughout the whole redshift range. This is easily done
by displaying the absolute magnitude and redshift, and cut the sample 
at an absolute magnitude $M_{V}<=-21$  using the SubSample
function (see figure \ref{tot_sample}b).
We save such selection in an ASCII file, for further use.\\
Now that we have a subsample clean from selection effects, we can go 
back to our original problem, and first of all we want to see whether
the effect of D4000 decreasing as redshift increases is still noticeable.
\begin{figure}
\begin{center}
\includegraphics[width=11.0cm]{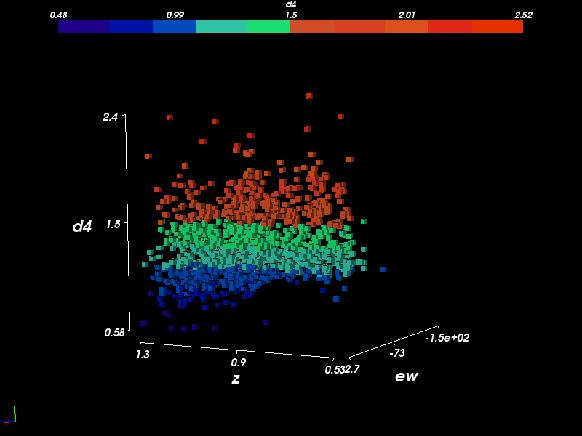}
\includegraphics[width=11.0cm]{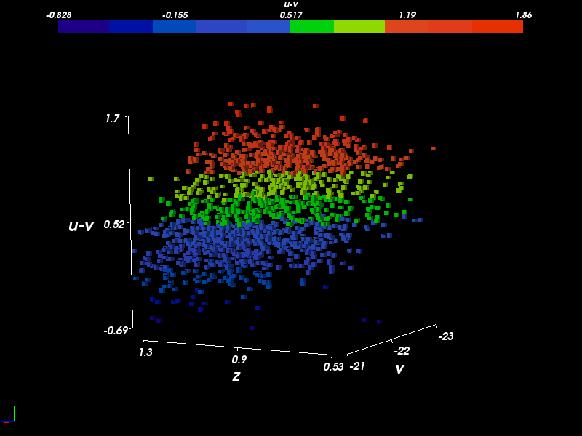}
\caption{Evolution of abundance of star forming galaxies with
  redshift: on the top we show the spectrophotometric indicators
  D4000 and [OII], on the bottom the photometric indicators U-V and
  $M_{V}$. In both cases, a trend of increased abundance of galaxies 
 showing low star
  age (top) and bluer colours (bottom) at higher redshifts is visible}
\label{starform_evol}
\end{center}
\end{figure} 

In fig. \ref{starform_evol}a, we show the D4000 vs. [OII] vs. z
datacube, where for clarity we use 4 different colours for different
D4000 ranges: at higher redshifts, not only
D4000 assumes smaller values, but rotating the cube we see that 
[OII] has a tendency to be higher (on average) with increasing
redshift. 
These two effects point towards a higher star forming rate at higher
redshift, and especially for $z>=1.1$. If we use the second
star-formation indicator, i.e. the colour magnitude diagram vs. redshift
(see fig. \ref{starform_evol} b), we see again that at higher redshift, blue
galaxies are more abundant than red galaxies.\\
As a last point, we want to see whether galaxies which would be defined as
quiescent in a colour magnitude diagram are also spectroscopically
quiescent
(i.e. galaxies belonging  to the red sequence should not show emission
lines), and vice-versa, if galaxies which show an old stellar
population
are indeed red in intrinsic colour. 
\begin{figure}
\begin{center}
\includegraphics[width=11cm]{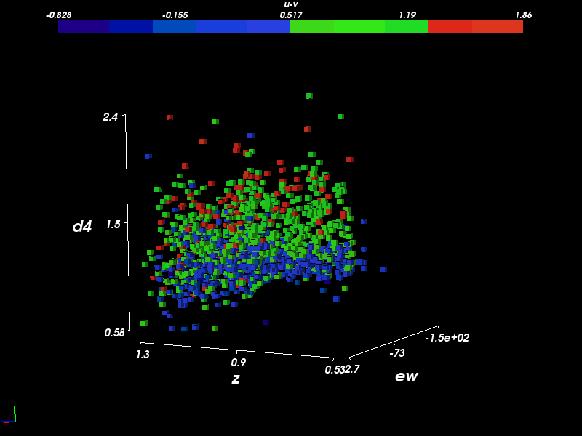}
\includegraphics[width=11cm]{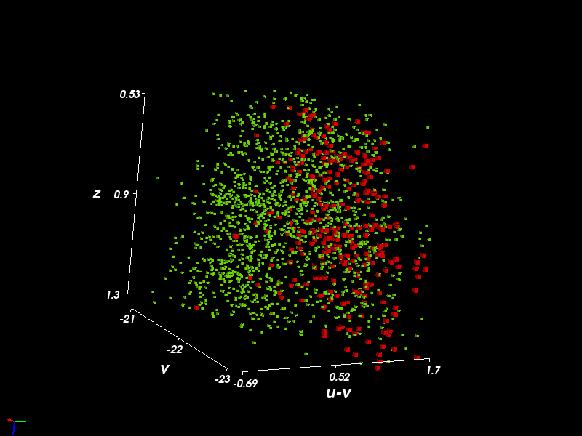}
\caption{Top (a): D4000 vs. [OII] vs. redshift data cube. The red sequence
galaxies are indicated as red, the bluest one in blue, while green
dots are galaxies in between. Bottom (b): colour magnitude diagram vs. redshift: 
red dots are now galaxies with old stellar age and no sign of star formation}
\label{starform_evol1}
\end{center}
\end{figure} 

Displaying again the D4000 vs. [OII] vs. redshift data cube
(fig. \ref{starform_evol1}a), we now
use colour to indicate the galaxy intrinsic U-V: as expected, red
galaxies show, on average, a higher stellar age indicator. Still,
there are some red galaxies showing a low value for D4000, as well as
blue galaxies with an old stellar population (high D4000), and
especially the intermediate class of colours (the green dots) span the
whole plane. This high degree of mixing does not seem to depend on
redshift. Also in the colour magnitude diagram
(fig. \ref{starform_evol1}b), this is clearly visible: we have now
highlighted in red the galaxies which show a D4000$>=1.5$ (old stellar
age) AND a low value of [OII] (no sign of star formation): these
galaxies have a tendency to have redder colours, but are not necessarily
the reddest ones, and not all the red
galaxies fall in this category\\
To summarize, using VisIVO exclusively, and without any programming
knowledge whatsoever, we have been able to inspect our sample and
1) see the volume selection and extract a bias free subsample,
2) find an interesting evolution of galaxy stellar age with redshift,
which deserves further statistical inspection \citep{ver07} for
a more exhaustive discussion of this topic),
3) show that selecting red galaxies on the basis of spectroscopic
features vs. photometric colours does not give the same sample. The
degree of contamination in the different cases is thoroughly
discussed in \citep{fra07}.

\subsection{Detecting Shocks in Galaxy Clusters} 

Cosmic shock waves are believed to be among the most efficient accelerators of particles 
in the universe. The final spectrum of
the accelerated particles is influenced by the complex interplay between the growth of
cosmic structures, the geometry of the shock waves and the number of shocks that a particle
may experience during its life.
A consistent and complete description of the dynamics of cosmological shock waves
is far from being achieved. However, numerical simulations provide a valuable  
contribution to the comprehension of this process.
We have performed a large number of simulations using
the Enzo code, which is an adaptive mesh refinement cosmological code
(http://cosmos.ucsd.edu/enzo), developed by Bryan et al.
\citep{bry95}. The Enzo code couples
an N-body particle-mesh solver with an PPM Eulerian adaptive method for ideal 
gas-dynamics by Colella \& Woodward 1984. 
For all the runs, we adopted the standard "Concordance" model, with
density parameters $\Omega_0 = 1.0$, $\Omega_{BM} = 0.044$, $\Omega_{DM} =
0.226$, $\Omega_{\Lambda} = 0.73$, Hubble parameter $h = 0.71$ and initial spectrum 
normalization $\sigma_8 = 0.94$.
In order to have a large cluster statistics and a wide cosmological volume,
we simulated  several datacubes, finally assembling an overall
volume of $\sim 100Mpc/h$ per side\\

\begin{figure}
\begin{center}
\includegraphics[width=9.0cm]{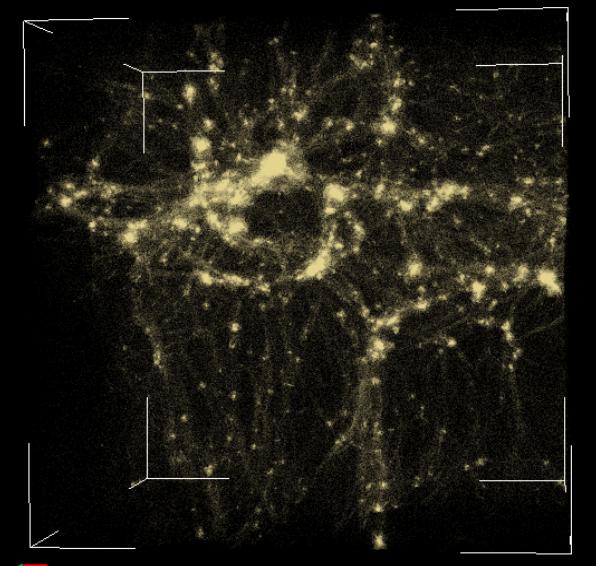}
\includegraphics[width=9.0cm]{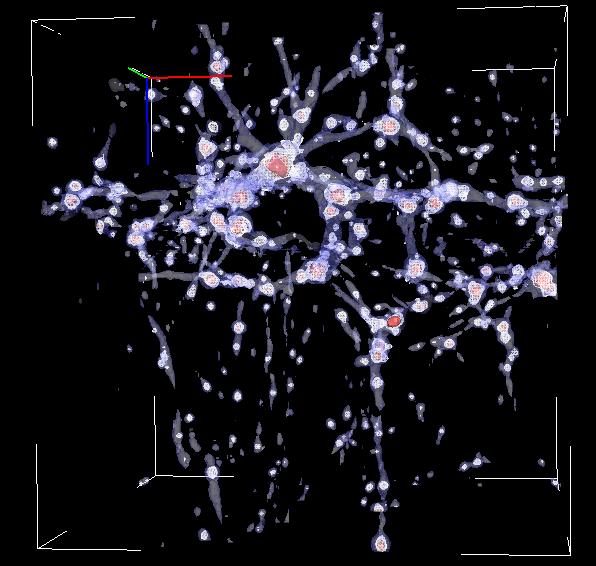}
\caption{The mass distribution of dark and baryonic matter in a $40^{3}$Mpc$^{3}$ cosmological
simulation. Dark matter particles and gas isosurfaces are compared. }
\label{bmdm}
\end{center}
\end{figure}

\begin{figure}
\begin{center}
\includegraphics[width=11.0cm]{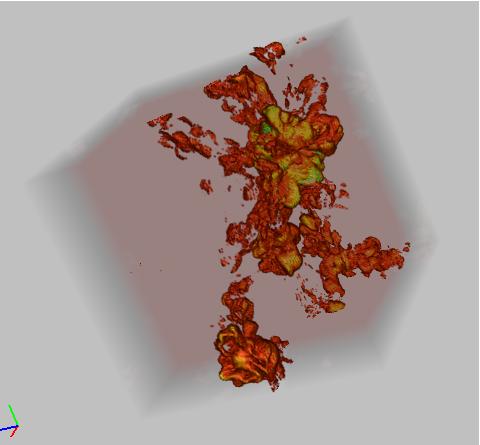}
\includegraphics[width=11.0cm]{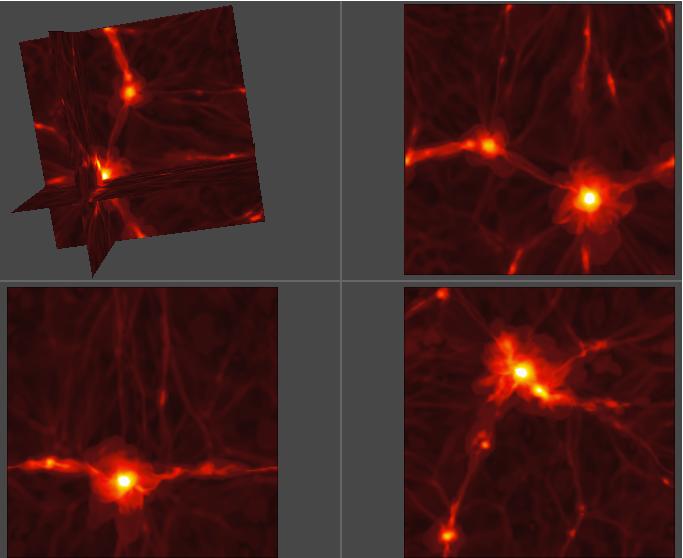}
\caption{The temperature distribution of the simulated gas (top panel) is
rendered using a Ray Tracing technique. X-ray maps (bottom panel) are
visualized using a Ortho Slice utility. }
\label{tempxray}
\end{center}
\end{figure}

Shocks identification has been performed using a novel method which studies the structure of the baryonic velocity
field and evaluates Mach numbers through velocity jumps, developed by Vazza et al. \citep{vaz07}.
Cells which present velocity jumps that can produce a  
shock wave \citep{lan59} are tagged as {\it shocked cells} and their  
Mach number is estimated as:

\begin{equation}
M=(3/(1-4\cdot \Delta v))^{1/2},
\label{eq:vel}
\end{equation}

where $\Delta v$ is the fluid velocity jump across the shock, in the  reference
frame of the shock itself, $\Delta v = v_{pre}-v_{post}$.
This technique, coupled with the unprecedented good resolution
of our data in the outermost regions of virialized structures (where shocks very often occur), 
allows us to study dynamical regimes never seen before. The use of a visualization code
such as VisIVO is of great importance because it allows us to detect 
and follow the behaviours of shock patterns which are characterized by very complex 
volume--filling properties, which had always been erased by most standard reduction analyses and
can be of primary importance to correctly describe the observational impact of these
mechanisms.\\
Data from the simulations are saved in raw binary and HDF5 formats, 
which generates a huge data collection.
In the following, for simplicity, we will focus on the results of one of the simulations,
characterized by $160^3$ mesh cells and the same number of N-Body particles.
The computational box is 40 Mpc and the presented results are at redshift $z=0.0$.
However, all the conclusions can be extended to all the outputs of our data 
collection. VisIVO is used to explore in an intuitive and effective way
the data and to focus on the interesting and, sometimes, unexpected aspects.
Data have been combined
using the Merge Tables function. The Math Op. tools have been used to derive 
new datasets, like logarithmic quantities or velocity magnitude.\\
Cosmological simulations produce extremely complex 
structures, characterized by clumps, filaments, sheets, large voids etc. In figure \ref{bmdm},
the distribution of particles representing the dark matter is shown in the top 
panel. The bottom panel presents a comparison between the dark matter and the gas mass 
distributions, these two quantities being represented by isosurfaces. Red and light blue isocontours show
the gas distribution at different mass density values. For comparison, dark matter isosurfaces 
(in white colour) are superimposed. Dark matter mass density is calculated by smoothing
the particle masses on the computational grid using the VisIVO point distribute function.
Gas and Dark matter mass distributions are similar, since they are both driven 
by a common gravitational potential. The temperature distribution, shown in the top panel figure
\ref{tempxray} using a volume rendering technique, follows the overall mass distribution.
However, a shock wave which forms during the gravitational collapse events expands rapidly, 
raising the gas temperature to $10^6$-$10^7$K on volumes up to $\sim 10^3$ Mpcs.
In order to have a quantity directly comparable to observations, we have
calculated the gas X-ray emission due to thermal bremsstrahlung 
directly using the VisIVO Math tool. The X-ray
distribution is shown using the VisIVO slider utility in the bottom panel of 
figure \ref{tempxray}. As expected the emission is strongly associated with the 
clusters, since it depends principally on the mass concentration.

\begin{figure}
\begin{center}
\includegraphics[width=11.0cm]{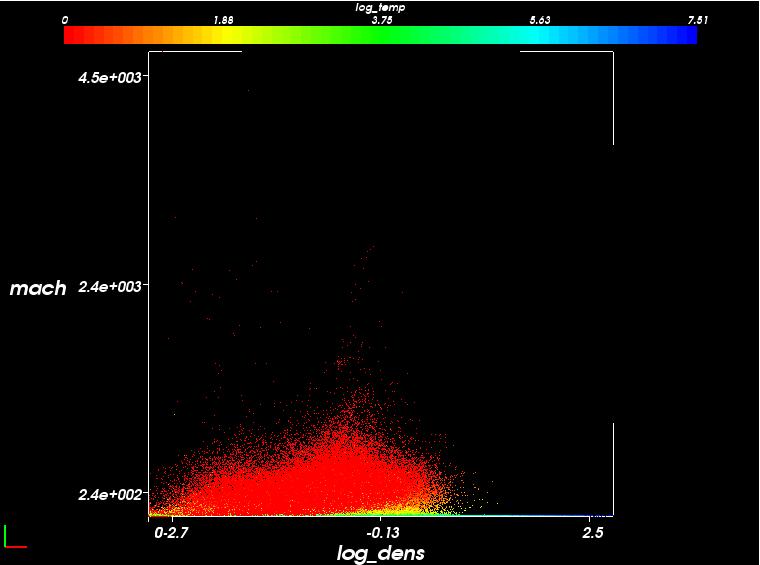}
\includegraphics[width=11.0cm]{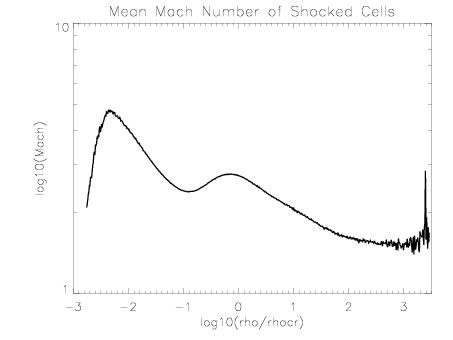}
\caption{The relation between the mach number of shocked regions and the mass density. 
The existence of two populations of shocks at the overdensities of $\rho/\rho_cr 
\sim 2\cdot 10^{-2}$ and $\rho/\rho_cr \sim 1$ is evident. In the top panel more than 4 millions
points from the analyzed simulation are displayed and coloured by their temperature. In the bottom panel the distribution
calculated using all the available datasets is shown.}
\label{rhomach2d}
\end{center}
\end{figure}

Once general features of the data have been analyzed, we 
switch to the identification and characterization of shocks. 
Shock fronts and the corresponding Mach numbers are identified by our 
{\it velocity-jumps} based procedure. 
The shocks-density distribution is presented in figure \ref{rhomach2d}.
\begin{figure}
\begin{center}
\includegraphics[width=11.0cm]{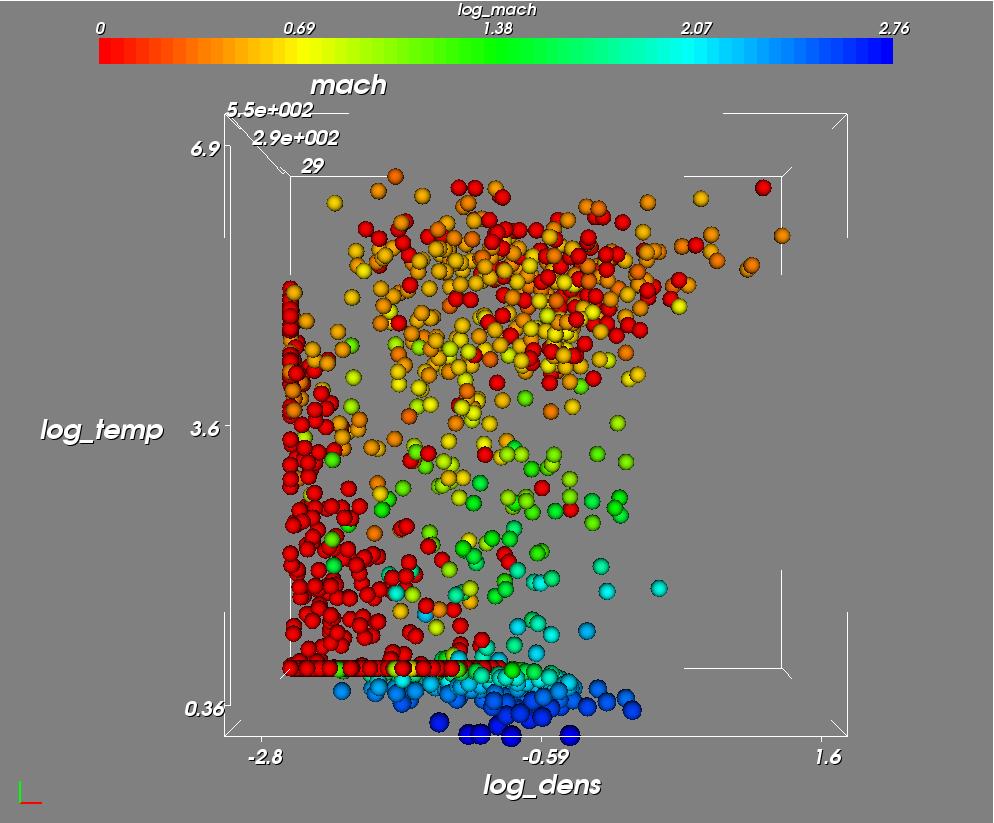}
\includegraphics[width=11.0cm]{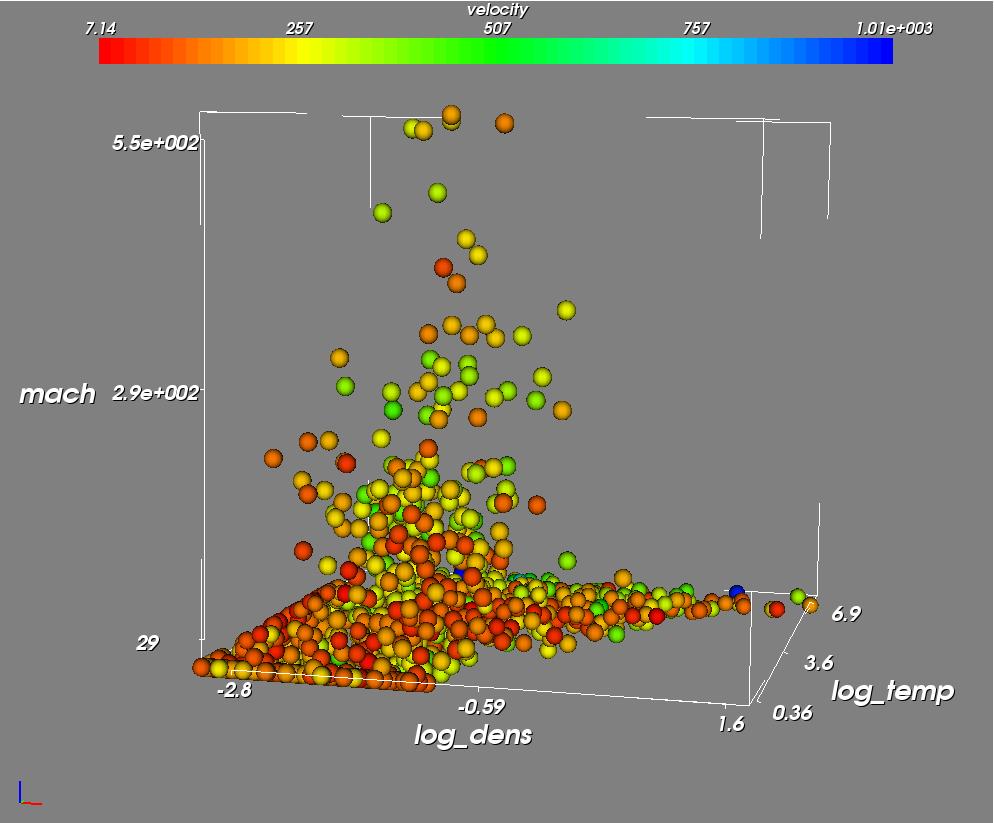}
\caption{Phase space distribution of a random subsample of the simulation 
the Cartesian axes are represented by mass density, temperature and Mach numbers. 
Colours are associated to the Mach numbers (top panel) and the velocity magnitude
(bottom panel). }
\label{mach}
\end{center}
\end{figure}

\begin{figure}
\begin{center}
\includegraphics[width=9.0cm]{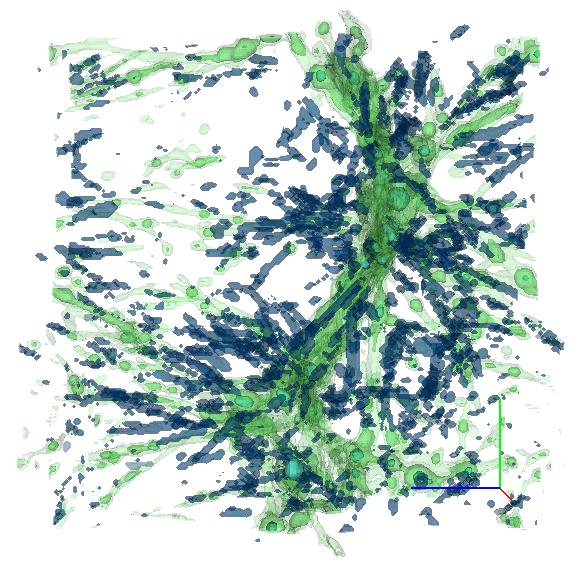}
\includegraphics[width=9.0cm]{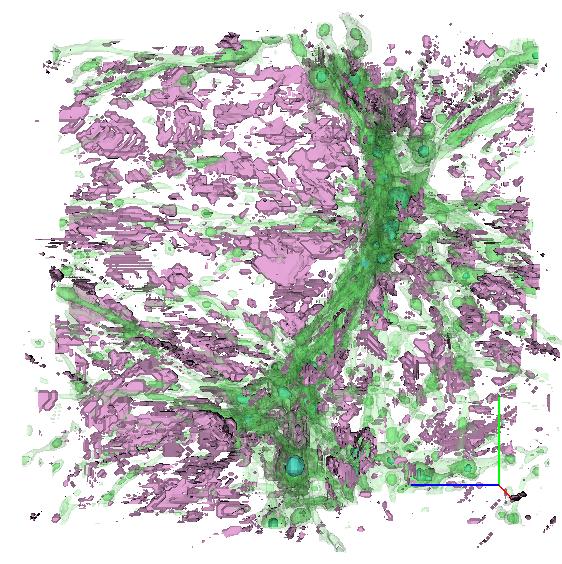}
\caption{Shocked regions for two different density values, corresponding to the Mach number peaks 
of figure \ref{rhomach2d}. The top panel shows the surface a Mach$>2.5$ for the right peak (blue surface). 
The bottom panel shows the surface at Mach$>2.5$ for the left peak (violet surface). For comparison 
a mass density isosurface is shown (green surface in both panels).}
\label{shocks}
\end{center}
\end{figure}

In the top panel, points are coloured according to their temperature. 
Most of the high Mach (Mach number $M > 10$) shocks are in low temperature regions, therefore
far outside the galaxy cluster virial radius. 
The distribution present two peaks corresponding to different values of
density. The right peak (larger density value) is observed in any kind of simulation
and it corresponds to shocks forming in the outer outskirts of clusters 
and along filaments, where the matter accretion is still strong.
The left peak corresponds to rare voids, which represent the low density tail of the
probability density function in a $\Lambda$CDM universe. It
has never been detected in any previous numerical simulations. The double peaked feature 
of the distribution is confirmed when all the dataset available in our collection, are taken into 
account, as shown in bottom panel of figure \ref{rhomach2d}.\\
This result is determined both by the improved accuracy of our 
shock reconstruction algorithm and by the simulations code. In fact 
some of the previous studies are related to SPH based simulation, which cannot
treat in a proper way the hydrodynamics of underdense regions, that are poorly
sampled by the particle based approach.
Figure \ref{mach} shows a 4D phase space distribution of a subsample
of points extracted from the whole simulation. The VisIVO Randomizer function
allows the user to select a small number of points in order to proceed 
faster in the analysis, without requiring huge computational 
resources. The figure shows the density-temperature-mach number distribution 
of sampled points, which are further coloured by the mach number itself (top panel)
and the velocity (bottom panel). The top panel shows that high mach points (blue
spheres) are concentrated at low temperature. High density regions also have  
high temperatures. However, low density regions, have temperatures in a wide
range, from $\sim 0$ to $\sim 10^7$K. This is due to the fast shock propagation 
outside collapsing structures (see also figures \ref{bmdm} and \ref{tempxray}).
The bottom panel emphasizes the density-mach number relation. It is still present,
even if less clear, due to lower statistics, the low density
Mach number peak. The L-shaped phase space distribution
(high mach-low temperatures, low mach-high temperatures) is also evident. The velocity 
seems not to present a specific trend with respect to the other quantities.
All these features 
are present also using the complete dataset and 
in any other random extraction, proving that there are not 
spurious effects connected to statistical biases.\\ 
In figure \ref{shocks}, we show the geometric distribution of the shocks, 
visualized as 3D surfaces with Mach $>2.5$ and selected according to the mass density
values. The top panel shows the shocks on the high density 
peak of the figure \ref{rhomach2d} distribution. The bottom panel shows the shocks
in low density regions. 
The presence of such vast surfaces of high mach shock in 
underdense regions suggests that these regions must be carefully taken into account, in order to properly 
evaluate the net amount of energy which ends up in particles re-acceleration \citep{bru03}.\\
A further analysis of the data is ongoing. Quantitative data analysis tools are
at this point necessary to estimate the results that have been revealed and
emphasized using VisIVO. The $visual-based$ method proved to be extremely 
effective in getting an immediate and effective approach to a complex and huge
dataset, selecting and extracting interesting features, which would otherwise be hardly 
detectable.

\begin{figure}
\begin{center}
\includegraphics[width=14.0cm]{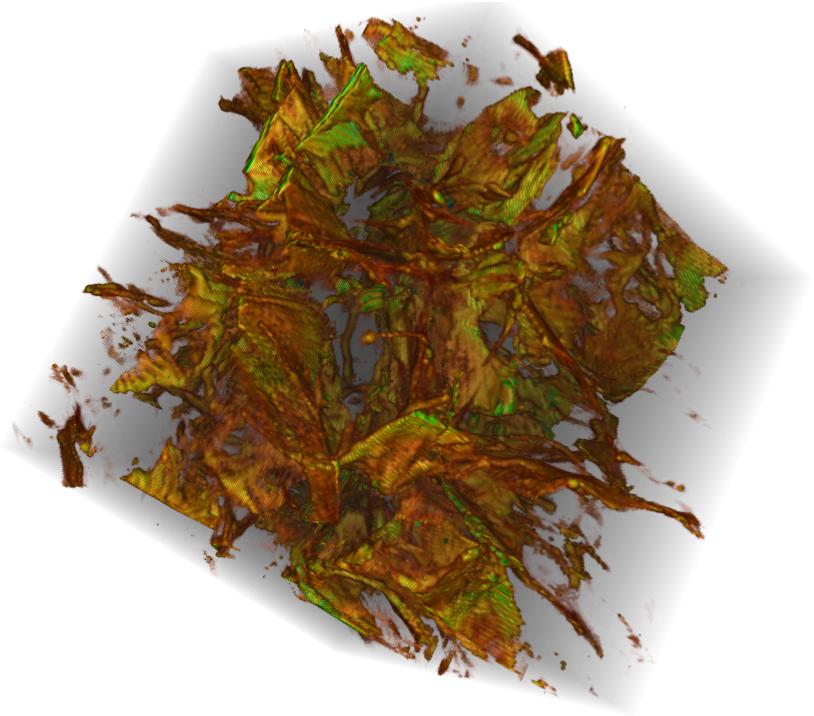}
\caption{The overall volume distribution of shocks. The Mach number increases from red to green to blue.}
\label{shocks2}
\end{center}
\end{figure}

\section{Summary}

In the previous sections we have shown how an advanced visualization tool
like VisIVO can be used for helping the researcher in analyzing complex data.
Visualization cannot provide quantitative results, but it allows the user 
to have an immediate and intuitive approach to the data. The various 3D rendering 
techniques supported by VisIVO, together with the possibility 
of visualizing complementary quantities with colours, glyphs and vectors, allows
the user to discriminate between data features at a glance, pointing out special characteristics
and focusing on interesting regions. The software also implements a limited but effective set of statistical
tools, that can be used to make quick estimates of the properties
of a distribution. Mathematics functions let the user derive new fields starting from
the original ones. VisIVO is being developed to follow the IVOA recommendations and standards,
so that it is interoperable with the Virtual Observatory framework. Furthermore, it
supports the PLASTIC protocol to allow the user to use the software 
together with other tools, such as Aladin or Topcat, which have  complementary data analysis capacities
to those of VisIVO. In this way the researcher can have a complete and customized cooperating
set of tools that makes his/her research activity more and more efficient and focused on scientific issues.

New features of VisIVO will focus on exploiting new hardware  
architectures that are rapidly appearing in many desktop
machines, such as 64 bit and multicore systems, subject to
making effective use of such capabilities in the tool.
The opportunity to use VisIVO in data centres and on dedicated
visualization servers will drive the new releases of the code.
We will also investigate the possibility of interacting with remote  
data services that support the SNAP protocol, which  is an emerging  
protocol for retrieving data from numerical simulations. Finally,  
VisIVO will be developed to integrate with the VO's Theoretical Data  
Archive framework.
VisIVO will display a subset of the whole data file, which will  
generally be very large, and will allow the user  to select a  
spherical or rectangular region and retrieve, through a remote service,
the extracted sub-sample.

\acknowledgments

We gratefully acknowledge useful discussion with Dr. F. Pasian and Dr. R. Smareglia from INAF Astronomical Observatory of Trieste,  
and we would like to thank F. Genova, F. Bonnarel and T. Boch from CDS - Observatoire Astronomique de Strasbourg, 
Dr. R. Mann  from the Institute for Astronomy,
University of Edinburgh  for the useful help they gave us,  Franco Vazza (Dep. of Astronomy--INAF-IRA Bologna) for the suggestions and the 
contributes given for the analysis of cosmological 
simulations part.  and Dr. L. Santagati of INAF - Catania Astrophysical Observatory for the  useful revision of the text. Numerical simulations
have been performed using the IBM-SP5 system at
CINECA, Bologna, with CPU time assigned under an INAF-CINECA agreement 2006/2007.




\clearpage

\end{document}